\documentclass[italian,english]{article}
\usepackage[latin9]{inputenc}
\usepackage{color}
\usepackage{amstext}
\usepackage{amssymb}

\makeatletter
\newcommand{\lyxaddress}[1]{
\par {\raggedright #1
\vspace{1.4em}
\noindent\par}
}

\makeatother

\usepackage{babel}

\begin{document}

\title{\textbf{Cosmology of Einstein-Vlasov system in a weak modification
of general relativity}}

\author{\textbf{Christian Corda}}

\maketitle

\lyxaddress{\begin{center}
International Institute for Theoretical Physics and Mathematics Einstein-Galilei,
via Santa Gonda 14, Prato Italy and Institute for Basic Research,
P. O. Box 1577, Palm Harbor, FL 34682, USA 
\par\end{center}}

\lyxaddress{\begin{center}
\textit{E-mail address:} \textcolor{blue}{cordac.galilei@gmail.com}
\par\end{center}}
\begin{abstract}
In earlier work it was shown that a weak modification of general relativity,
in the linearized approach, renders a spherically symmetric and stationary
model of the Universe. This was due to the presence of a third mode
of polarization in the linearized gravity in which a {}``\emph{curvature
energy}'' term is present. Such term was identified as the Dark Energy
of the Universe. 

In this letter a more realistic model is discussed. A different cosmological
solution to the Einstein-Vlasov System is analysed. This solution
shows reasonable results which are within the standard bounds predicted
by the cosmological observations. 
\end{abstract}

Over the past century, a standard cosmological model has emerged.
With relatively few parameters, the model describes the evolution
of the Universe and astronomical observations on scales ranging from
a few to thousands of Megaparsecs. In this model the Universe is spatially
flat, homogeneous and isotropic on large scales, composed of radiation,
ordinary matter (electrons, protons, neutrons and neutrinos), non-baryon
Cold Dark Matter, and Dark Energy. The galaxies and the large-scale
structures grew gravitationally from tiny, nearly scale-invariant
adiabatic Gaussian fluctuations \cite{key-1,key-2}. The WMAP data
offers a demanding quantitative test of this model \cite{key-3}.
In other words, the Universe is seen like a dynamic and thermodynamic
system. The test masses (i.e. the {}``particles'') are the galaxies
which are stellar systems with a number of the order of $10^{9}-10^{11}$
stars. The galaxies are located in clusters and super clusters, and
observations show that, on cosmological scales, their distribution
is uniform. These assumptions can be summarized in the Cosmological
Principle: \textit{the Universe is homogeneous everywhere and isotropic
around every point} \cite{key-1,key-2}. The Cosmological Principle
simplifies the analysis of the large scale structure. It implies that
the proper distances between any two galaxies is given by a universal
scale factor which is the same for any couple of galaxies \cite{key-1,key-2}. 

On the other hand, although the standard model of the Universe achieved
great success, it also displayed many shortcomings and flaws. As a
matter of fact, the accelerated expansion of the Universe, which is
observed today, implies that cosmological dynamics is dominated by
the so called Dark Energy, which gives a large negative pressure.
This is the standard picture, in which this new ingredient is considered
as a source on the right-hand side of the field equations. It should
be some form of un-clustered, non-zero vacuum energy which, together
with the clustered Dark Matter, drives the global dynamics. This is
the so called \textquotedblleft{}concordance model\textquotedblright{}
($\Lambda$CDM) which gives, in agreement with the CMBR, LSS and SNeIa
data, a good picture of the observed Universe today, but presents
several shortcomings such as the well known \textquotedblleft{}coincidence\textquotedblright{}
and \textquotedblleft{}cosmological constant\textquotedblright{} problems
\cite{key-4}. An alternative approach is changing the left-hand side
of the field equations, to see if the observed cosmic dynamics can
be achieved by extending General Relativity \cite{key-5}. In this
different context, it is not required to find candidates for Dark
Energy and Dark Matter, that, till now, have not been found. Only
the \textquotedblleft{}observed\textquotedblright{} ingredients, which
are curvature and baryon matter, have to be taken into account. Considering
this point of view, one can think that gravity is different at various
scales and there is room for alternative theories. In principle, the
most popular Dark Energy and Dark Matter models can be achieved considering
$f(R)$ Theories of Gravity, where $R$ is the spacetime curvature,
see the review \cite{key-5}, the recent results \cite{key-6}-\cite{key-10}
and references within. In this picture, the sensitive detectors for
gravitational waves (GWs), like bars and interferometers, whose data
analysis recently started \cite{key-11}, could, in principle, be
important. In fact, a consistent GW astronomy will be the definitive
test for General Relativity or, alternatively, a strong endorsement
for Extended Theories of Gravity \cite{key-12}.

In earlier work \cite{key-13} it was shown that a weak modification
of General Relativity, in the linearized approach, renders a spherically
symmetric and stationary model of the Universe. This was due to the
presence of a third mode of polarization in the linearized gravity
in which a {}``\emph{curvature energy}'' term is present. Such term
was identified as the Dark Energy of the Universe. 

In this letter a more realistic model is discussed. A different cosmological
solution to the Einstein-Vlasov System is analysed. This solution
shows reasonable results which are within the standard bounds predicted
by the cosmological observations. 

Let us consider the action\begin{equation}
S=\int d^{4}x\sqrt{-g}f_{0}R^{1+\varepsilon}+\mathcal{L}_{m}\label{eq: high order e basta}\end{equation}

This is a particular choice in $f(R)$ Theories of Gravity \cite{key-5}-\cite{key-10}. 

In the limit $\varepsilon\rightarrow0$ and $f_{0}=1$ the action
(\ref{eq: high order e basta}) recovers the canonical form of the
Einstein-Hilbert action of General Relativity \cite{key-1,key-2},
i.e. 

\begin{equation}
S=\int d^{4}x\sqrt{-g}R+\mathcal{L}_{m}\label{eq: standard}\end{equation}

Criticisms on $f(R)$ Theories of Gravity arise from the fact that
lots of such theories can be excluded by requirements of Cosmology
and Solar System tests \cite{key-14}. However, in the case of the
action (\ref{eq: high order e basta}), the discrepancy with respect
to standard General Relativity is very weak, because $\varepsilon$
is a very small real parameter. Thus, the mentioned constraints could,
in principle, be satisfied. In particular the authors of \cite{key-14}
found 

\begin{equation}
0\leq\varepsilon\leq7.2*10^{-19}.\label{eq: minimum constrain}\end{equation}
Fundamental constrains can be renormalized in order to obtain $f_{0}=1$
in the action (\ref{eq: high order e basta})

We analyse interactions at cosmological scales, therefore the linearized
theory in vacuum, i.e. with $\mathcal{L}_{m}=0$, can be considered.
It gives a better approximation than the Newtonian theory \cite{key-1,key-2,key-13}.
Thus, let us start from the pure curvature action \begin{equation}
S=\int d^{4}x\sqrt{-g}f_{0}R^{1+\varepsilon}.\label{eq: high order 12}\end{equation}

The theory arising from such an action has been linearized in \cite{key-13}.
A short review of the linearized approach is needed for a better understanding
of the theoretical framework.

By varying the action (\ref{eq: high order 12}) with respect to $g_{\mu\nu},$
the field equations are obtained (through this letter the convention
$G=1$, $c=1$ and $\hbar=1$ will be used and Greek indices run from
0 to 3)) \cite{key-13} \begin{equation}
G_{\mu\nu}=\frac{1}{(1+\varepsilon)f_{0}R^{\varepsilon}}\{-\frac{1}{2}g_{\mu\nu}\varepsilon f_{0}R^{1+\varepsilon}+[(1+\varepsilon)f_{0}R^{\varepsilon}]_{;\mu;\nu}-g_{\mu\nu}\square[(1+\varepsilon)f_{0}R^{\varepsilon}]\}.\label{eq: einstein 2}\end{equation}

The trace of these field equations is 

\begin{equation}
\square(1+\varepsilon)f_{0}R^{\varepsilon}=\frac{(1-\varepsilon)}{3}f_{0}R^{1+\varepsilon}.\label{eq: KG}\end{equation}

By making the identifications \cite{key-13}

\begin{equation}
\begin{array}{ccccc}
\Phi\rightarrow(1+\varepsilon)f_{0}R^{\varepsilon} &  & \textrm{and } &  & \frac{dV}{d\Phi}\rightarrow\frac{(1-\varepsilon)}{3}f_{0}R^{1+\varepsilon}\end{array}\label{eq: identifica}\end{equation}

a Klein - Gordon equation for the effective $\Phi$ scalar field is
obtained. It can be written as

\begin{equation}
\square\Phi=\frac{dV}{d\Phi}.\label{eq: KG2}\end{equation}

Let us consider a little perturbation of the background, which is
assumed given by a near Minkowskian background, i.e. a Minkowskian
background plus $\Phi=\Phi_{0}$ (the Ricci scalar is assumed constant
in the background) \cite{key-13}. We also assume $\Phi_{0}$ to be
a minimum for the effective potential $V$: 

\begin{equation}
V\simeq\frac{1}{2}\alpha\delta\Phi^{2}\Rightarrow\frac{dV}{d\Phi}\simeq E^{2}\delta\Phi,\label{eq: minimo}\end{equation}

and the constant $E$ has mass-energy dimension. 

Let us write

\begin{equation}
\begin{array}{c}
g_{\mu\nu}=\eta_{\mu\nu}+h_{\mu\nu}\\
\\\Phi=\Phi_{0}+\delta\Phi.\end{array}\label{eq: linearizza}\end{equation}

To first order in $h_{\mu\nu}$ and $\delta\Phi$, calling $\widetilde{R}_{\mu\nu\rho\sigma}$
, $\widetilde{R}_{\mu\nu}$ and $\widetilde{R}$ the linearized quantity
which correspond to $R_{\mu\nu\rho\sigma}$ , $R_{\mu\nu}$ and $R$,
the linearized field equations are \cite{key-13}

\begin{equation}
\begin{array}{c}
\widetilde{R}_{\mu\nu}-\frac{\widetilde{R}}{2}\eta_{\mu\nu}=(\partial_{\mu}\partial_{\nu}h_{\varepsilon}-\eta_{\mu\nu}\square h_{\varepsilon})\\
\\{}\square h_{\varepsilon}=E^{2}h_{\varepsilon},\end{array}\label{eq: linearizzate1}\end{equation}

where 

\begin{equation}
h_{\varepsilon}\equiv\frac{\delta\Phi}{\Phi_{0}}.\label{eq: definizione}\end{equation}

Notice that even if the minimum of the effective potential $V$ is
the real number $0,$ $V$ is not equal to $\Phi,$ therefore we are
not dividing by $0$ in Eq. (\ref{eq: definizione}). In other words,
as $V\neq\Phi,$ the condition $V=0$ does\emph{ }not imply $\Phi_{0}=0.$ 

Then, from the second of Eqs. (\ref{eq: linearizzate1}), one can
define the mass-energy like \cite{key-13}

\begin{equation}
E\equiv\sqrt{\frac{\square h_{\varepsilon}}{h_{\varepsilon}}}=\sqrt{\frac{\square\delta\Phi}{\delta\Phi}}=\sqrt{\frac{\square\delta R^{\varepsilon}}{\delta R^{\varepsilon}}}.\label{eq: massa}\end{equation}

Thus, as the mass-energy is generated by variation of the Ricci scalar,
we can say that, in a certain sense, it is generated by variation
of spacetime curvature. This is exactly the {}``\emph{curvature energy}''
term \cite{key-13}. 

We recall that, in the present case, the theory is viable as the modification
of General Relativity is very weak and in agreement with requirements
of Cosmology and Solar System tests \cite{key-14}. 

In \cite{key-13} it has been shown that, by using gauge transformations,
the linearization process gives

\begin{equation}
h_{\mu\nu}(t,z)=A^{+}(t-z)e_{\mu\nu}^{(+)}+A^{\times}(t-z)e_{\mu\nu}^{(\times)}+h_{\varepsilon}(t,z)\eta_{\mu\nu}.\label{eq: perturbazione totale}\end{equation}

The term $A^{+}(t-z)e_{\mu\nu}^{(+)}+A^{\times}(t-z)e_{\mu\nu}^{(\times)}$
describes the two standard polarizations of gravitational waves which
arise from General Relativity, while the term $h_{\varepsilon}(t,z)\eta_{\mu\nu}$
is the massive field arising from the high order theory. In other
words, the function $R^{\varepsilon}$ of the Ricci scalar generates
a third polarization state for GWs which is not present in standard
General Relativity. This third polarization has a {}``curvature''
energy $E,$ see \cite{key-13} for details.

In the simple model of Universe in \cite{key-13} the dynamic of the
matter is described by the Vlasov equation \cite{key-13,key-15}.
The gravitational forces between the particles, i.e., the galaxies,
are supposed to be mediated by the third mode of Eq. (\ref{eq: perturbazione totale})
after making the assumption that at cosmological scales such a mode
becomes dominant (i.e. $A^{+},A^{-}\ll h_{\varepsilon}$) \cite{key-13}.
In this way the {}``curvature'' mass-energy $E$ can be identified
as the Dark Energy of the Universe $\simeq10^{-29}g/cm^{3}$ \cite{key-3}. 

The model in \cite{key-13} realized a spherically symmetric and stationary
model of the Universe. The present work analyses a different cosmological
solution to the Einstein-Vlasov System. The new model shows reasonable
results which are within the standard bounds predicted by the cosmological
observations. 

In the hypothesis $A^{+},A^{-}\ll h_{\varepsilon}$, Eq. (\ref{eq: perturbazione totale})
can be rewritten as

\begin{equation}
h_{\mu\nu}(t,z)=h_{\varepsilon}(t,z)\eta_{\mu\nu},\label{eq: perturbazione scalare}\end{equation}
and the line element of the model is the conformally flat one 

\begin{equation}
ds^{2}=[1+h_{\varepsilon}(t,z)](dt^{2}-dz^{2}-dx^{2}-dy^{2}).\label{eq: metrica puramente scalare}\end{equation}
By defining 

\begin{equation}
a^{2}\equiv1+h_{\varepsilon}(t,z),\label{eq: a quadro}\end{equation}

the line element (\ref{eq: metrica puramente scalare}) results similar
to the cosmological Friedmann-Robertson-Walker (FRW) line element
of the standard homogeneous, isotropic and flat Universe \cite{key-1,key-2}

\begin{equation}
ds^{2}=[a^{2}(t,z)](+dt^{2}-dz^{2}-dx^{2}-dy^{2}).\label{eq: metrica FRW}\end{equation}

Strictly speaking, this metric does not describe an homogeneous and
isotropic universe, as $h_{\varepsilon}$ is also a function of z.
Thus, we have to further assume $\partial_{z}h_{\varepsilon}=0$,
which removes the z-dependence \cite{key-16}.

To satisfy the condition demanding that the particles make up an ensemble
with no collisions in the spacetime, the particle density must be
a solution of the Vlasov equation

\begin{equation}
\partial_{t}f+\frac{p^{a}}{p^{0}}\partial_{x^{a}}f-\Gamma_{\mu\nu}^{a}\frac{p^{\mu}p^{\nu}}{p^{0}}\partial_{p^{a}}f=0.\label{eq: Vlasov}\end{equation}
Here $\Gamma_{\mu\nu}^{\alpha}$ represent the usual connections,
$f$ is the particle density and $p^{0}$ is determined by $p^{a}$($a=1,2,3$)
according to the relation \begin{equation}
g_{\mu\nu}p^{\mu}p^{\nu}=-1\label{eq: mass-shell}\end{equation}
\cite{key-13,key-15}, which expresses the condition that the four
momentum $p^{\mu}$ lies on the mass shell of the metric. 

A clarification is needed. We have to distinguish between the standard
kinetic theory which is based upon Boltzmann equation \cite{key-17}
and the Einstein\textendash{}Vlasov system which we are discussing
in the present work. In a kinetic description the time evolution of
the system is determined by the interactions between the particles,
which depend on the physical situation \cite{key-18}. For instance,
the driving mechanism for the time evolution of a neutral gas is the
collision between particles (the Boltzmann equation) \cite{key-18}.
For a plasma the interaction is through the electromagnetic field
produced by the charges (the Vlasov\textendash{}Maxwell system), and
in astrophysics the interaction between collisionless particles is
gravitational (the Vlasov\textendash{}Poisson system and the Einstein\textendash{}Vlasov
system) \cite{key-18}. Therefore, an important difference between
the standard kinetic theory and the Einstein\textendash{}Vlasov system
which we are discussing in this letter is that in the first case there
are collisions between particles while in the Einstein\textendash{}Vlasov
system particles interact gravitationally only and they are \emph{collisionless}.

We recall that, in general, the Vlasov- Poisson system is \cite{key-13,key-15}

\begin{equation}
\begin{array}{c}
\partial_{t}f+v\cdot\bigtriangledown_{x}f-\bigtriangledown_{x}U\cdot\bigtriangledown_{v}f=0\\
\\\bigtriangleup U=4\pi\rho\\
\\\rho(t,x)=\int dvf(t,x,v),\end{array}\label{eq: VP}\end{equation}

where $t$ denotes the time and $x$ and $v$ the position and the
velocity of the galaxies. The function $U=U(t,x)$ is the average
Newtonian potential generated by the galaxies. This system represents
the non-relativistic kinetic model for an ensemble of particles with
no collisions, which interacts through the gravitational forces that
they generate collectively \cite{key-13,key-15}. Thus, one can use
such a system to describe the motion of galaxies within the Universe,
thought of as pointlike particles, when the relativistic effects are
negligible \cite{key-13,key-15}. In this approach, the function $f(t,x,v)$
in the Vlasov- Poisson system (\ref{eq: VP}) is non-negative and
gives the density on phase space of the galaxies within the Universe.

By using the classical transformation from conformal time to synchronous
time \cite{key-1} 

\begin{equation}
dt\rightarrow\frac{dt}{\sqrt{1+h_{\varepsilon}(t)}}\label{eq: trasformazione temporale}\end{equation}

the line element (\ref{eq: metrica puramente scalare}), in spherical
coordinates, becomes \begin{equation}
ds^{2}=dt^{2}-[1+h_{\varepsilon}(t)](dr^{2}+r^{2}(d\theta^{2}+\sin^{2}\theta d\varphi^{2})).\label{eq: metrica puramente scalare piena 2}\end{equation}

The metric tensor has the form

\begin{equation}
g_{\mu\nu}=\left(\begin{array}{ccc}
1 &  & 0\\
\\0 &  & \gamma_{mn}\end{array}\right),\label{eq: tensore metrico}\end{equation}

where $\gamma_{mn}=1+h_{\varepsilon}(t).$

Following \cite{key-2}, we define $\chi_{mn}\equiv\frac{\partial}{\partial t}\gamma_{mn.}$
The Einstein field equations in the synchronous frame are: \begin{equation}
R_{0}^{0}=-\frac{1}{2}\frac{\partial}{\partial t}\chi_{a}^{a}-\frac{1}{4}\chi_{a}^{b}\chi_{b}^{a}=(T_{0}^{0}-\frac{1}{2}T)\label{eq: sincrona 1}\end{equation}

\begin{equation}
R_{a}^{0}=\frac{1}{2}(\chi_{a;b}^{b}-\chi_{b;a}^{a})=T_{a}^{0}\label{eq: sincrona 2}\end{equation}

\begin{equation}
R_{a}^{b}=-P_{a}^{b}-\frac{1}{2\sqrt{\gamma}}\frac{\partial}{\partial t}(\sqrt{\gamma}\chi_{a}^{b})=T_{a}^{b}-\frac{1}{2}\delta_{a}^{b}T,\label{eq: sincrona 3}\end{equation}

where $P_{a}^{b}$ is the Ricci tensor in $3$ dimensions \cite{key-2}.

On the other hand, the Einstein field equations in the Einstein-Vlasov
system are \cite{key-13,key-15}

\begin{equation}
G_{\mu\nu}=\frac{2}{\sqrt{-g}}\int f(t,x,p)p_{\mu}p_{\nu}\delta(p^{2}+m^{2})d^{4}p,\label{eq: G}\end{equation}

where $m$ is the mass of a particle (galaxy).

We can split the function $f(t,x,p)$ into a couple of equations for
$f_{+}(t,x,p)$ and $f_{-}(t,x,p)$ which are constructed by reducing
$f(t,x,p)$ respectively on the \textquotedblright{}upper\textquotedblright{}
half and on the \textquotedblright{}lower\textquotedblright{} half
of the mass shell. Eq. (\ref{eq: Vlasov}) becomes

\begin{equation}
\partial_{t}f_{\pm}=-\frac{1}{p_{\pm}^{0}}\left(\gamma^{mn}p_{n}\frac{\partial}{\partial x_{m}}-\frac{1}{2}\frac{\partial\gamma^{nr}}{\partial x_{m}}p_{n}p_{r}\frac{\partial}{\partial p_{m}}\right)f_{\pm}.\label{eq: Vlasov 2}\end{equation}

Eq. (\ref{eq: Vlasov 2}) can be interpreted in Hamiltonian terms:

\begin{equation}
p_{\pm}^{0}\partial_{t}f_{\pm}=\{H,f_{\pm}\},\label{eq: interpretazione hamiltoniana}\end{equation}

where the Hamiltonian function is 

\begin{equation}
H\equiv\frac{1}{2}\gamma^{mn}p_{m}p_{n}.\label{eq: hamiltoniana}\end{equation}

One can calculate the components of energy-momentum tensor $T_{\mu\nu}$
in the approximation which considers galaxies like massless particles
($m=0$ in Eq. (\ref{eq: G}))

\begin{equation}
T_{00}=\frac{1}{(\sqrt{1+h_{\varepsilon}(t)})^{3}r^{2}\sin\theta}\int\frac{f_{+}+f_{-}}{\sqrt{1+h_{\varepsilon}(t)}}\sqrt{\frac{p_{1}^{2}+p_{2}^{2}}{r^{2}}+\frac{p_{3}^{2}}{r^{2}\sin\theta}}d^{3}p\label{eq: zero-zero}\end{equation}

\begin{equation}
T_{mn}=\frac{1}{(\sqrt{1+h_{\varepsilon}(t)})^{3}r^{2}\sin\theta}\int\sqrt{1+h_{\varepsilon}(t)}\frac{(f_{+}+f_{-})}{\sqrt{\frac{p_{1}^{2}+p_{2}^{2}}{r^{2}}+\frac{p_{3}^{2}}{r^{2}\sin\theta}}}p_{m}p_{n}d^{3}p\label{eq: m-n}\end{equation}

\begin{equation}
T_{0m}=\frac{1}{(\sqrt{1+h_{\varepsilon}(t)})^{3}r^{2}\sin^{2}\theta}\int(f_{+}-f_{-})p_{m}d^{3}p.\label{eq: zero-m}\end{equation}

The Einstein field equations (\ref{eq: sincrona 1}), (\ref{eq: sincrona 2})
and (\ref{eq: sincrona 3}) give two independent dynamic equations
which can be written down in terms of the scale factor $a=\sqrt{1+h_{\varepsilon}(t)}$:

\begin{equation}
\dot{a}^{2}=-1+\frac{1}{3a}\int(f_{+}(s)+f_{-}(s))\frac{s}{a}d^{3}s\label{eq: s1}\end{equation}

\begin{equation}
\ddot{a}=-\frac{2}{a}-2\frac{\dot{a}^{2}}{a}+\frac{1}{a^{2}}\int(f_{+}(s)+f_{-}(s))d^{3}s,\label{eq: s2}\end{equation}

where \begin{equation}
s\equiv p_{1}^{2}+\frac{p_{2}^{2}}{r^{2}}+\frac{p_{3}^{2}}{r^{2}\sin^{2}\theta}.\label{eq: s}\end{equation}

By introducing the dimensionless variables $\underline{r}$ and $\underline{t}$
we put

\begin{equation}
\begin{array}{c}
a=a_{0}\underline{r}\\
\\t=a_{0}\underline{t}\\
\\\dot{\underline{r}}=\frac{d\underline{r}}{d\underline{t}}\\
\\j=\frac{1}{3}a_{0}^{2}\rho_{0},\end{array}\label{eq: new variables}\end{equation}

where $a_{0}$ is the present value of the scale factor of the Universe.

Eq. (\ref{eq: s1}) becomes \begin{equation}
\begin{array}{c}
\dot{\underline{r}}^{2}=-1+\frac{j}{\underline{r}^{2}}\\
\\\underline{r}_{0}=1.\end{array}\label{eq: system}\end{equation}

The solution of the system (\ref{eq: system}) is

\begin{equation}
\underline{r}(\underline{t})=\sqrt{j-(\underline{t}-\sqrt{j-1})^{2}}\label{eq: solution}\end{equation}

if $j\geq1.$ Notice that expression (\ref{eq: solution}) becomes
imaginary for certain values of the parameter $\underline{t}$ \cite{key-17}.
Such a parameter represents the cosmic time normalized by the present
value of the scale factor $a_{0}$ , see Eq. (\ref{eq: new variables}).
By choosing the origin of the cosmic time at the present era of the
cosmological evolution, i.e. $t_{0}=0,$ expression (\ref{eq: solution})
is \emph{always} real if $j\geq1.$ 

Returning to the ($t,$ $a$) variables we get:

\begin{equation}
a(t)=a_{0}\sqrt{\frac{a_{0}^{2}\rho_{0}}{3}-(\frac{t}{a_{0}}-\sqrt{\frac{a_{0}^{2}\rho_{0}}{3}}-1)^{2}}.\label{eq: soluzione}\end{equation}

The today's observed Hubble radius and the today's observed density
of the Universe are \foreignlanguage{italian}{respectively \cite{key-3}
$a_{0}\gtrsim10^{28}{\normalcolor cm}$ and $\rho_{0}\approx10^{-57}{\normalcolor cm^{-2}}$.
Therefore $j\approx1.$}

By inserting these data in Eq. (\ref{eq: soluzione}) we obtain a
singularity at a time

$t_{s}\approx-10^{10}{\normalcolor seconds}$ and a value for the
today's theoretical Hubble constant $H_{0}=\frac{\dot{a}_{0}}{a_{0}}\approx10^{-29}{\normalcolor cm^{-1}}.$

As it follows from the above analysis, even under the assumption to
neglect the baryon mass of the galaxies the results look reasonable.
They are of the same order of magnitude of the standard cosmological
model \foreignlanguage{italian}{\cite{key-3}}.

\subsubsection*{Conclusion remarks}

In earlier work it was shown that a weak modification of general relativity,
in the linearized approach, renders a spherically symmetric and stationary
model of the Universe. This was due to the presence of a third mode
of polarization in the linearized gravity in which a {}``\emph{curvature
energy}'' term is present. Such term was identified as the Dark Energy
of the Universe. 

In this letter a more realistic model has been discussed. A different
cosmological solution to the Einstein-Vlasov System has been analysed.
This solution shows reasonable results which are within the standard
bounds predicted by the cosmological observations. 

We hope in further analyses which could insert the baryon mass too
and realize a better \emph{fine-tuning} the model with the cosmological
observations.

\subsubsection*{Acknowledgements}

The Ruggero M. Santilli Foundation has to be thanked for supporting
this letter (Grant Number RMS-TH-5735A2310).

I strongly thank an unknown referee for useful comments and advices
which permitted to improve this letter.

\end{document}